\documentstyle[12pt]{article}
\begin{document}
\title {ANISOTROPIC GEODESIC FLUID SPHERES IN GENERAL RELATIVITY.}
\author{L. Herrera\thanks{Postal address: Apartado 80793, Caracas 1080 A,
Venezuela; e-mail address:
laherrera@telcet.net.ve}
\\
Escuela de F\'\i sica, Facultad de Ciencias,\\
Universidad Central de Venezuela,\\
Caracas, Venezuela.\\
\\
{J. Martin\thanks{e-mail address:chmm@usal.es}} and J. Ospino\\
Area de F\'\i
sica Te\'orica. Facultad de Ciencias.\\ Universidad de Salamanca. 37008
Salamanca, Espa\~na.
}
\date{}
\maketitle

\begin{abstract}
It is shown that unlike the perfect fluid case, anisotropic fluids
(principal stresses unequal) may be geodesic, without this implying the
vanishing of (spatial) pressure gradients.
Then the condition of vanishing four acceleration is integrated in
non-comoving coordinates. The resulting models are necessarily dynamic, and
the mass
function is expressed in terms of the fluid velocity as measured by a
locally Minkowskian observer. An explicit example is worked out.
\end{abstract}

\newpage
\section{Introduction}

As is well known, the vanishing of four acceleration (geodesic condition)
implies for a perfect fluid that pressure gradients vanish. In the case of
spherical bounded ( non-dissipative)
configuration, the vanishing of pressure at the boundary surface, implies
in turn the vanishing of pressure everywhere within the distribution
(dust).

Indeed, for a perfect fluid the equation of motion reads
\begin{equation}
(\rho + p) a^\alpha = h^{\alpha\nu} p_{,\nu}
\label{em}
\end{equation}
with
\begin{equation}
h^\alpha_\mu \equiv \delta^\alpha_\mu - u^\alpha u_\mu
\label{h}
\end{equation}
\begin{equation}
a^\mu = u^\nu u^\mu_{;\nu}
\label{a}
\end{equation}
where colon and semicolon denote partial and covariant derivatives, and as
usual $a^\mu$, $u^\mu$, $\rho$ and $p$ stand for the four acceleration, the
four velocity, the energy density
and the pressure respectively.

>From the above it becomes evident that the geodesic condition implies the
vanishing of pressure gradients. From purely physics considerations this
conclusion  is also obvious: the
vanishing of four-acceleration means that only gravitational forces are
acting on any fluid element, thereby implying that pressure gradients (the
only hydrodynamic force in a perfect
fluid) vanish. However in the case of anisotropic fluids, an additional
force term appears besides the pressure gradient (see next section).
Therefore it is in principle possible to
have a fluid distribution, such that both terms cancel each other, leading
to a geodesic fluid with non-vanishing pressure gradients.

Since the original Lemaitre paper \cite{Lem} and particularly since the
work of Bowers and Liang \cite{BL} anisotropic fluids have attracted the
attention of many
researchers in relativity and relativistic astrophysics (see \cite{Anis}
and references therein), due to the conspicuous role played by local
anisotropy of pressure in the structure
and evolution of self--gravitating objects. It is the purpose of this work
to present further models of anisotropic spheres, based on the geodesic
condition. Besides the natural
interest of such models in general relativity, the presented models are
interesting because they represent the generalization of Tolman--Bondi
\cite{ToB} models to anisotropic fluids,
in non-comoving coordinates. Incidentally, it is worth noticing that in the
classical paper by Oppenheimer and Snyder on dust collapse \cite{Opp}, they
start their study, using the same
kind of coordinates we use here, and then switch to comoving ones, in order
to integrate the field equations.

The plan of the paper is as follows. In Section 2 we define the conventions
and give the field equations and expressions for the kinematic variables we
shall need, in
noncomoving coordinates. The geodesic condition is explicitly integrated
in Section 3. In Section 4 we work out an example. Finally a discussion of
results
is presented in Section 5.

\section{Relevant  Equations and Conventions}
We consider spherically symmetric distributions of collapsing anisotropic
fluid, which  we assume to evolve adiabatically (without dissipation),
bounded by a
spherical surface $\Sigma$.

\noindent
The line element is given in Schwarzschild--like coordinates by

\begin{equation}
ds^2=e^{\nu} dt^2 - e^{\lambda} dr^2 -
r^2 \left( d\theta^2 + sin^2\theta d\phi^2 \right),
\label{metric}
\end{equation}

\noindent
where $\nu(t,r)$ and $\lambda(t,r)$ are functions of their arguments. We
number the coordinates: $x^0=t; \, x^1=r; \, x^2=\theta; \, x^3=\phi$.

\noindent
The metric (\ref{metric}) has to satisfy Einstein field equations

\begin{equation}
G^\nu_\mu=-8\pi T^\nu_\mu,
\label{Efeq}
\end{equation}

\noindent
which in our case read \cite{Bo}:

\begin{equation}
-8\pi T^0_0=-\frac{1}{r^2}+e^{-\lambda}
\left(\frac{1}{r^2}-\frac{\lambda'}{r} \right),
\label{feq00}
\end{equation}

\begin{equation}
-8\pi T^1_1=-\frac{1}{r^2}+e^{-\lambda}
\left(\frac{1}{r^2}+\frac{\nu'}{r}\right),
\label{feq11}
\end{equation}

\begin{eqnarray}
-8\pi T^2_2  =  -  8\pi T^3_3 = & - &\frac{e^{-\nu}}{4}\left(2\ddot\lambda+
\dot\lambda(\dot\lambda-\dot\nu)\right) \nonumber \\
& + & \frac{e^{-\lambda}}{4}
\left(2\nu''+\nu'^2 -
\lambda'\nu' + 2\frac{\nu' - \lambda'}{r}\right),
\label{feq2233}
\end{eqnarray}

\begin{equation}
-8\pi T_{01}=-\frac{\dot\lambda}{r},
\label{feq01}
\end{equation}

\noindent
where dots and primes stand for partial differentiation with respect
to $t$ and $r$,
respectively.

\noindent
In order to give physical significance to the $T^{\mu}_{\nu}$ components
we apply the Bondi approach \cite{Bo}.

\noindent
Thus, following Bondi, let us introduce purely locally Minkowski
coordinates ($\tau, x, y, z$) (alternatively one may introduce a tetrad
field associated to locally Minkowskian observers).

$$d\tau=e^{\nu/2}dt\,;\qquad\,dx=e^{\lambda/2}dr\,;\qquad\,
dy=rd\theta\,;\qquad\, dz=rsin\theta d\phi.$$

\noindent
Then, denoting the Minkowski components of the energy tensor by a bar,
we have

$$\bar T^0_0=T^0_0\,;\qquad\,
\bar T^1_1=T^1_1\,;\qquad\,\bar T^2_2=T^2_2\,;\qquad\,
\bar T^3_3=T^3_3\,;\qquad\,\bar T_{01}=e^{-(\nu+\lambda)/2}T_{01}.$$

\noindent
Next, we suppose that when viewed by an observer moving relative to these
coordinates with proper velocity $\omega(t,r)$ in the radial direction, the
physical
content  of space consists of an anisotropic fluid of energy density $\rho$,
radial pressure $P_r$ and  tangential pressure $P_\bot$. Thus, when viewed
by this moving
observer the covariant tensor in
Minkowski coordinates is

\[ \left(\begin{array}{cccc}
\rho     &  0  &   0     &   0    \\
0 &  P_r     &   0     &   0    \\
0       &   0       & P_\bot  &   0    \\
0       &   0       &   0     &   P_\bot
\end{array} \right). \]

\noindent
Then a Lorentz transformation readily shows that

\begin{equation}
T^0_0=\bar T^0_0= \frac{\rho + P_r \omega^2 }{1 - \omega^2} ,
\label{T00}
\end{equation}

\begin{equation}
T^1_1=\bar T^1_1=-\frac{ P_r + \rho \omega^2}{1 - \omega^2},
\label{T11}
\end{equation}

\begin{equation}
T^2_2=T^3_3=\bar T^2_2=\bar T^3_3=-P_\bot,
\label{T2233}
\end{equation}

\begin{equation}
T_{01}=e^{(\nu + \lambda)/2} \bar T_{01}=
-\frac{(\rho + P_r) \omega e^{(\nu + \lambda)/2}}{1 - \omega^2},
\label{T01}
\end{equation}
\noindent

Note that the coordinate velocity in the ($t,r,\theta,\phi$) system, $dr/dt$,
is related to $\omega$ by

\begin{equation}
\omega(t,r)=\frac{dr}{dt}\,e^{(\lambda-\nu)/2}.
\label{omega}
\end{equation}

\noindent
Feeding back (\ref{T00}--\ref{T01}) into (\ref{feq00}--\ref{feq01}), we get
the field equations in  the form

\begin{equation}
\frac{\rho + P_r \omega^2 }{1 - \omega^2}
=-\frac{1}{8 \pi}\Biggl\{-\frac{1}{r^2}+e^{-\lambda}
\left(\frac{1}{r^2}-\frac{\lambda'}{r} \right)\Biggr\},
\label{fieq00}
\end{equation}

\begin{equation}
\frac{ P_r + \rho \omega^2}{1 - \omega^2}=-\frac{1}{8
\pi}\Biggl\{\frac{1}{r^2} - e^{-\lambda}
\left(\frac{1}{r^2}+\frac{\nu'}{r}\right)\Biggr\},
\label{fieq11}
\end{equation}

\begin{eqnarray}
P_\bot = -\frac{1}{8 \pi}\Biggl\{\frac{e^{-\nu}}{4}\left(2\ddot\lambda+
\dot\lambda(\dot\lambda-\dot\nu)\right) \nonumber \\
 - \frac{e^{-\lambda}}{4}
\left(2\nu''+\nu'^2 -
\lambda'\nu' + 2\frac{\nu' - \lambda'}{r}\right)\Biggr\},
\label{fieq2233}
\end{eqnarray}

\begin{equation}
\frac{(\rho + P_r) \omega e^{(\nu + \lambda)/2}}{1 -
\omega^2}=-\frac{\dot\lambda}{8 \pi r}.
\label{fieq01}
\end{equation}

\noindent
At the outside of the fluid distribution, the spacetime is that of
Schwarzschild,
given by

\begin{equation}
ds^2= \left(1-\frac{2M}{r}\right) dt^2 - \left(1-\frac{2M}{r}\right)^{-1}dr^2 -
r^2 \left(d\theta^2 + sin^2\theta d\phi^2 \right),
\label{Vaidya}
\end{equation}

\noindent
As is well known, in order to match smoothly the two metrics above on the
boundary surface
$r=r_\Sigma(t)$, we must require the continuity of the first and the second
fundamental
form across that surface. In our notation this implies

\begin{equation}
e^{\nu_\Sigma}=1-\frac{2M}{r_\Sigma},
\label{enusigma}
\end{equation}
\begin{equation}
e^{-\lambda_\Sigma}=1-\frac{2M}{r_\Sigma}.
\label{elambdasigma}
\end{equation}
and
\begin{equation}
\left[P_r\right]_\Sigma=0,
\label{PQ}
\end{equation}
Where, from now on, subscript $\Sigma$ indicates that the quantity is
evaluated at the boundary surface $\Sigma$.

Eqs. (\ref{enusigma}), (\ref{elambdasigma}) and (\ref{PQ}) are the necessary and
sufficient conditions for a smooth matching of the two metrics (\ref{metric})
and (\ref{Vaidya}) on $\Sigma$.

\noindent
Next, let us write the energy momentum tensor in the form
\begin{equation}
T_{\mu\nu} = \left(\rho+P_\bot\right)u_\mu u_\nu - P_\bot g_{\mu\nu} +
\left(P_r-P_\bot\right)s_\mu s_\nu
\label{T-}
\end{equation}

with
\begin{equation}
u^\mu=\left(\frac{e^{-\nu/2}}{\left(1-\omega^2\right)^{1/2}},\,
\frac{\omega\, e^{-\lambda/2}}{\left(1-\omega^2\right)^{1/2}},\,0,\,0\right),
\label{umu}
\end{equation}
\begin{equation}
s^\mu=\left(\frac{\omega \, e^{-\nu/2}}{\left(1-\omega^2\right)^{1/2}},\,
\frac{e^{-\lambda/2}}{\left(1-\omega^2\right)^{1/2}},\,0,\,0\right),
\label{smu}
\end{equation}

where $u^\mu$ denotes the four velocity of the fluid and  $s^\mu$ is a
radially directed space--like vector orthogonal to $u^\mu$. Then the radial
component of the
conservation law

\begin{equation}
T^\mu_{\nu;\mu}=0.
\label{dTmn}
\end{equation}
may be written as

\begin{equation}
\left(-8\pi T^1_1\right)'=\frac{16\pi}{r} \left(T^1_1-T^2_2\right)
+ 4\pi \nu' \left(T^1_1-T^0_0\right) +
\frac{e^{-\nu}}{r} \left(\ddot\lambda + \frac{\dot\lambda^2}{2}
- \frac{\dot\lambda \dot\nu}{2}\right),
\label{T1p}
\end{equation}

\noindent
which in the static case becomes

\begin{equation}
P'_r=-\frac{\nu'}{2}\left(\rho+P_r\right)+
\frac{2\left(P_\bot-P_r\right)}{r},
\label{Prp}
\end{equation}

\noindent
representing the generalization of the Tolman--Oppenheimer--Volkof equation
for anisotropic fluids \cite{BL}. Thus, as mentioned before, local
anisotropy introduces an extra term in this ``force'' equation, besides the
usual pressure gradient term.

Finally, for the two non--vanishing components of the four acceleration, we
easily find

\begin{equation}
a_0=\frac{1}{1-\omega^2}\left[\left(\frac{\omega\dot\omega}{1-\omega^2} +
\frac{\omega^2 \dot\lambda}{2}\right) +
e^{\nu/2} e^{-\lambda/2}
\left(\frac{\omega \nu'}{2} +
\frac{\omega^2 \omega'}{1-\omega^2}\right)
\right],
\label{a0}
\end{equation}

\begin{equation}
a_1=-\frac{1}{1-\omega^2}\left[\left(\frac{\omega \omega'}{1-\omega^2} +
\frac{\nu'}{2}\right) +
e^{-\nu/2} e^{\lambda/2}
\left(\frac{\omega \dot\lambda}{2} +
\frac{\dot\omega}{1-\omega^2}\right)
\right],
\label{a1}
\end{equation}

\section{Integrating the Geodesic Condition}
Let us now integrate the geodesic condition. First, observe that from the
field equations (\ref{fieq00}),(\ref{fieq11}) and (\ref{fieq01}), one
obtains after simple manipulations
\begin{equation}
\omega e^{(\nu - \lambda)/2}(\lambda'+\nu') + (1+\omega^2) \dot\lambda=0
\label{gc1}
\end{equation}

Next, it follows at once from (\ref{a0}) and (\ref{a1}) that,
\begin{equation}
\omega a_1=-a_0 e^{(\lambda - \nu)/2}
\label{gc2}
\end{equation}

Therefore the vanishing four--acceleration condition amounts to
\begin{equation}
\left(\frac{\omega \omega'}{1-\omega^2} +
\frac{\nu'}{2}\right) +
e^{-\nu/2} e^{\lambda/2}
\left(\frac{\omega \dot\lambda}{2} +
\frac{\dot\omega}{1-\omega^2}\right)=0.
\label{gc3}
\end{equation}

Then, replacing $\nu'$ by its expression from (\ref{gc1}), into
(\ref{gc3}), this last equation becomes
\begin{equation}
\omega e^{(\nu - \lambda)/2}(\lambda'-\frac{2 \omega \omega'}{1-\omega^2})
+  \dot\lambda-\frac{2\omega \dot\omega}{1-\omega^2}=0,
\label{gc4}
\end{equation}
or, using (\ref{omega})
\begin{equation}
\dot\phi dt+\phi' dr=0,
\label{gc5}
\end{equation}
whose solution is
\begin{equation}
\phi=ln(1-\omega^2)+\lambda=Constant.
\label{gc6}
\end{equation}

Finally, from the fact that $\omega(t,0)=0$ we obtain,
\begin{equation}
e^{-\lambda}=1-\omega^2.
\label{gc7}
\end{equation}
Introducing the mass function as usually,
\begin{equation}
e^{-\lambda} = 1 - \frac{2m}{r}
\label{mass}
\end{equation}
we have
\begin{equation}
m = \frac{\omega^2 r}{2}
\label{3}
\end{equation}

In all the above we have assumed  $\omega\not =0$, since from simple
physical considerations we should not expect static solutions to exist.

Indeed, if we assume staticity ($\omega=0$) then the geodesic condition
implies $\nu'=0$, which in turn, using (\ref{fieq11}) and
(\ref{mass}),leads to
\begin{equation}
8\pi P_r=-\frac{2m}{r^3}.
\label{nost}
\end{equation}
Then junction condition (\ref{PQ}) would lead to $m_\Sigma=M=0$.

There is however one possible case of static geodesic solution, which
appears if we relax the condition of continuity of the second fundamental
form (implying the continuity of radial
pressure) across the boundary surface, and assume the existence of a
surface layer \cite{israel}.

In this specific case, it follows from (40), the geodesic condition and
field equations (15) and (17) that
\begin{equation}
\rho+ P_r+2 P_\bot=0
\label{rhopr}
\end{equation}

implying that the active gravitational mass (Tolman,\cite{Tol}) defined for
any $r<r_{\Sigma}$ as,

\begin{equation}
m_T = 4 \pi \int^{r}_{0}{r^2 e^{(\nu+\lambda)/2}
(T^0_0 - T^1_1 - 2 T^2_2) dr}
\label{Tol}
\end{equation}

vanishes inside the sphere.

We shall not consider here these kind of solutions and accordingly  all our
models will be dynamic ($\omega\neq 0$) and satisfy all junction conditions
.

Now, from (\ref{omega}) evaluated at the boundary surface, and
(\ref{enusigma}), (\ref{elambdasigma}), we obtain

\begin{equation}
\omega_\Sigma = \frac{\dot r_\Sigma}{1 - 2M/r_\Sigma}
\label{4}
\end{equation}

On the other hand, from (\ref{3}) evaluated at the boundary surface, we have

\begin{equation}
\omega_\Sigma = \pm \sqrt{\frac{2M}{r_\Sigma}}
\label{7}
\end{equation}
Where the $+$ ($-$) refers to the expansion (contraction) of the surface
(from now on we shall only consider the contracting case). Feeding back
(\ref{7}) into (\ref{4}), we get
\begin{equation}
\omega_\Sigma = \frac{\dot r_\Sigma}{1 - \omega_\Sigma^2}
\label{5}
\end{equation}

Then equating (\ref{5}) and (\ref{7}) we have
\begin{equation}
\dot r_\Sigma =  \left(\frac{2
M}{ r_\Sigma }\right)^{3/2}-\left(\frac{2
M}{ r_\Sigma }\right)^{1/2}
\label{11}
\end{equation}
This equation may be integrated to give
\begin{equation}
\frac{t}{2M} =  2
tanh^{-1}{\sqrt{\frac{2M}{r_\Sigma}}}-\frac{2 \left[1 + 6M/r_\Sigma\right]}{3
\left(2M/r_\Sigma\right)^{3/2}}
\label{12}
\end{equation}
giving the evolution of the boundary surface. Unfortunately this last
equation cannot be inverted (at least we were unable to do that) to obtain
the explicit form
$r=r_\Sigma(t)$.
Accordingly we have also integrated (\ref{11}) numerically , in order to
exhibit the evolution of $r_{\Sigma}$, see figure(1).

So far we have found all consequences derived from the geodesic condition
which, obviously, are valid in the pure dust case as well as in the
anisotropic case.  In the next section we
shall work out an explicit example by impossing an ``equation of state''
for the physical variables.

\section{A Model}
The purpose of this section is not to model any specific physical system,
but just to illustrate the consequences derived from the geodesic
condition.
Thus, somehow inspired by the incompressible fluid model, let us assume
\begin{equation}
T^0_0 = f(t)
\label{13}
\end{equation}
then from (\ref{3}) and the fact that
\begin{equation}
m'=4 \pi r^2 T^0_0
\label{14}
\end{equation}
one obtains
\begin{equation}
\omega =- r \sqrt{\frac{8 \pi}{3} f(t)}
\label{14}
\end{equation}
where
\begin{equation}
f(t) = \frac{3 M }{4 \pi r_\Sigma^3}.
\label{15}
\end{equation}
Observe  from (\ref{14}) that the evolution in this model is homologous.

Next, introducing the dimensionless variables
\begin{equation}
x \equiv \frac{r}{r_\Sigma} \qquad ; \qquad
y \equiv \frac{r_\Sigma}{2 M}
\label{16}
\end{equation}
we have
\begin{equation}
e^{-\lambda} = 1 - \frac{x^2}{y}
\label{17}
\end{equation}
\begin{equation}
\omega = -\frac{x}{\sqrt y}
\label{18}
\end{equation}
Finally, from the field equations the following relations follow
\begin{equation}
\rho y + P_r x^2 = \frac{3 (y - x^2)}{32 \pi M^2 y^3}
\label{19}
\end{equation}
\begin{equation}
P_r y + \rho x^2 = \frac{(y - x^2)}{8 \pi} \left[
\left(1 - \frac{x^2}{y}\right)
\left(\frac{1}{4 M^2 x^2 y^2} + \frac{\partial \nu}{\partial x}
\frac{1}{4 M^2 x y^2}\right) - \frac{1}{4 M^2 x^2 y^2}
\right]
\label{20}
\end{equation}
\begin{equation}
\rho + P_r = - \frac{3 \dot y (y - x^2)^{1/2} e^{-\nu/2}}{8 \pi (2 M y^3)}
\label{21}
\end{equation}

Then from (\ref{19}) and (\ref{20}) we obtain $\rho + P_r$ as function of
$\frac{\partial \nu}{\partial x}$, $x$ and $y$. Feeding back this
expression into (\ref{21}),this equation may be solved for $\nu$, which in
turn allows to express all physical
variables ($\rho, P_r$ and $P_\bot$) in terms of $x$ and $y$ which are
given by (\ref{12}) or alternatively by the numerical solution of
(\ref{11}).

\section{Conclusions}
We have seen that the geodesic condition, which  for anisotropic fluids  is
compatible with the presence of pressure gradients, can be integrated,
giving the explicit form  of the
evolution of the boundary surface. Resulting models may be regarded as
generalizations of Tolman--Bondi solutions, to anisotropic fluids. In order
to obtain the evolution of all
physical variables for different pieces of matter, additional information
has to be given. In the model above we have assumed condition (\ref{13}),
which in turn leads to the homology
condition (\ref{14}). Parenthetically, this last condition is widely used
by astrophysicists in their modelling of stellar structure and evolution
\cite{KW}.

Figures (1) and (2) display the behaviour of the radius and the evolution of
$\omega_\Sigma$ in the contracting case. As expected, as the boundary
surface approaches the horizon, its coordinate velocity ($\dot r_\Sigma$)
stalls, whereas the velocity
$\omega_\Sigma$ measured by the locally Minkowskian observer, tends to
light velocity. Figure (3) shows the sensitivity of the pattern evolution
with respect to the compactness of the
initial configuration. As expected more compact configurations collapse
faster. The good behaviour of
$\rho$ and
$p_r$ is easily deduced from (\ref{19})--(\ref{21}).

\section{Acknowledgements}
We acknowledge financial assistance under grant
BFM2000-1322 (M.C.T. Spain) and from C\'atedra-FONACIT, under grant
2001001789.

\newpage
\section{Figure captions}
\begin{itemize}
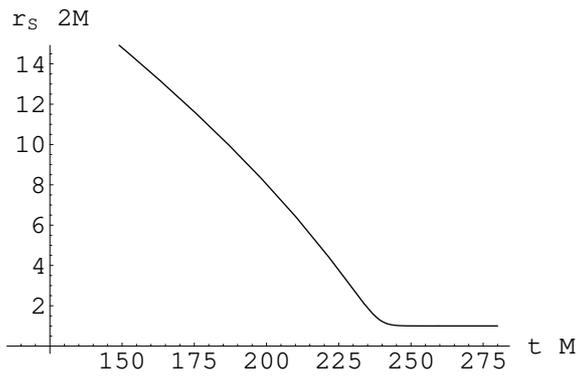
\item Figure 1. $y=r_\Sigma/2M$ as function of $t/M$ for the inital value
$y(0)=30$.

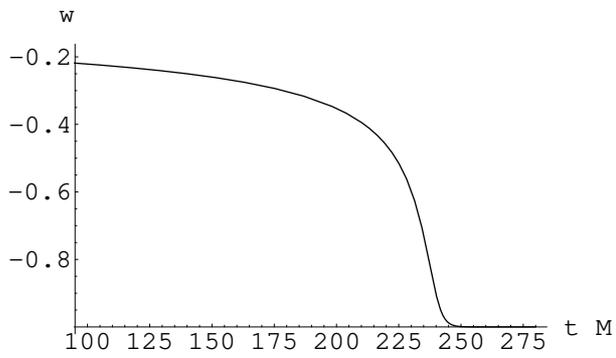
\item Figure 2. $\omega_\Sigma$ as function of $t/M$ for the same initial
data as in figure 1.

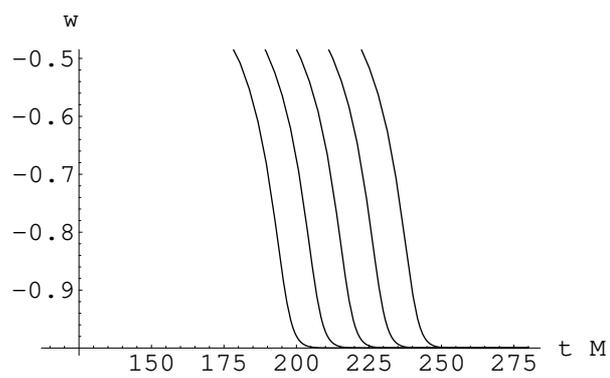
\item Figure 3. $\omega_\Sigma$ as function of $t/M$ for
$y(0)=30,29,28,27,26$ curves from rigth to left respectively.
\end{itemize}
\end{document}